\documentclass[twocolumn,prl,floatfix,superscriptaddress,showpacs,citeautoscript]{revtex4}
\usepackage{graphicx}
\usepackage{amssymb}
\usepackage{pifont}

\begin{document}

\title{Quantifying spin Hall angles from spin pumping: Experiments and Theory}

\author{O.~Mosendz}
\email{mosendz@anl.gov} \affiliation{Materials Science Division,
Argonne National Laboratory, Argonne, IL 60439, USA}

\author{J.~E.~Pearson}
\affiliation{Materials Science Division, Argonne National
Laboratory, Argonne, IL 60439, USA}

\author{F.~Y.~Fradin}
\affiliation{Materials Science Division, Argonne National
Laboratory, Argonne, IL 60439, USA}

\author{G.~E.~W.~Bauer}
\affiliation{Kavli Institute of NanoScience, Delft University of
Technology, 2628 CJ Delft, The Netherlands}

\author{S.~D.~Bader}
\affiliation{Materials Science Division, Argonne National
Laboratory, Argonne, IL 60439, USA} \affiliation{Center for
Nanoscale Materials, Argonne National Laboratory, Argonne, IL 60439,
USA}

\author{A.~Hoffmann}
\affiliation{Materials Science Division, Argonne National
Laboratory, Argonne, IL 60439, USA} \affiliation{Center for
Nanoscale Materials, Argonne National Laboratory, Argonne, IL 60439,
USA}



\date{\today}

\begin{abstract}
Spin Hall effects intermix spin and charge currents even in
nonmagnetic materials and, therefore, ultimately may allow the use
of spin transport without the need for ferromagnets.  We show how
spin Hall effects can be quantified by integrating
Ni$_{80}$Fe$_{20}|$normal metal (N) bilayers into a coplanar
waveguide.  A {\em dc} spin current in N can be generated by spin
pumping in a controllable way by ferromagnetic resonance. The
transverse {\em dc} voltage detected along the Ni$_{80}$Fe$_{20}|$N
has contributions from both the anisotropic magnetoresistance (AMR)
and the spin Hall effect, which can be distinguished by their
symmetries. We developed a theory that accounts for both.  In this
way, we determine the spin Hall angle quantitatively for Pt, Au and
Mo.  This approach can readily be adapted to any conducting material
with even very small spin Hall angles.
\end{abstract}



\maketitle


The spin-orbit interaction gives rise to spin-dependent scattering
that can couple charge and spin currents in conducting materials.
Resultant spin Hall effects
\cite{Dyakonov-JETF,Hirsch,Zhang-PRL2000} may therefore display
spin-dependent transport even in materials and device structures
that do not contain ferromagnetic materials.  The effectiveness of
this spin-charge conversion can be quantified by the
material-specific spin Hall angle $\gamma$, which is given by the
ratio of spin Hall and charge conductivities
\cite{Dyakonov-BookChapter2008} and can be quantified by
magnetotransport measurements
\cite{Fert-JMMM1981,Valenzuela-Nature2006,Kimura-PRL2007,Seki-NM2008,Morota-JAP09}.
However, $\gamma$ values reported in the literature vary over
several orders of magnitude even for nominally identical materials
(i.e., Au: $\gamma < 0.022$ \cite{Mihajlovic-Preprint} and $\gamma =
0.113$ \cite{Seki-NM2008}, and Pt: $\gamma = 0.0037$
\cite{Kimura-PRL2007} and $\gamma = 0.08$ \cite{Ando-PRL2008}). In
order to better understand spin-dependent scattering and its
potential use for spin-transport applications it is therefore highly
desirable to find a robust method to quantify $\gamma$.  Here we
demonstrate an approach whose sensitivity can be adjusted to measure
even very small values of $\gamma$.

Previous work on ferromagnetic resonance (FMR) in magnetic
multilayers has shown that spin pumping can create pure spin
currents in normal metals (N)
\cite{Heinrich-PRL2003,Woltersdorf-PRL2007,Mosendz-PRB09}.  Upon
excitation of FMR, the time-varying magnetization inside the
ferromagnet (F) generates an instantaneous spin current $j_s$ at the
F$|$N interface given by \cite{Brataas-PRL2002,Yaroslav-review}:
\begin{equation}\label{spincurrent}
j_s\vec{\rm{s}}=\frac{\hbar}{8\pi}Re
(2g_{\uparrow\downarrow})\left[\vec{\rm{m}}\times \frac{\partial
\vec{\rm{m}}}{\partial t}\right],
\end{equation}
where $\vec{\rm{m}}$ is the unit vector of the magnetization,
$\vec{\rm{s}}$ is the unit vector of the spin current polarization,
and $Re(g_{\uparrow\downarrow})$ is the real part of the spin mixing
conductance. Under a simple precession \cite{endnote} the spin
pumping induces a net {\em dc} spin current and by time-averaging
Eq.~(\ref{spincurrent}) we get:
\begin{equation}\label{dcspincurrent}
j_{s,dc}^{0}=\frac{\hbar\omega}{4\pi}Re
g_{\uparrow\downarrow}\sin^{2}\theta,
\end{equation}
where $\omega$ is the driving frequency and $\theta$ is the cone
angle of the precession of $\vec{\rm{m}}$.  This spin current decays
due to spin relaxation and diffusion in N, such that the spin
current at distance $z$ from the interface is:
\begin{equation}\label{zdepspincurrent}
j_{s,dc}(z) = j_{s,dc}^{0}\frac{\sinh
((z-t_{N})/\lambda_{sd})}{\sinh (t_{N}/\lambda_{sd})}
\end{equation}
where $\lambda_{sd}$ is the materials specific spin diffusion length
and $t_{N}$ is the thickness of the N layer.

The spin current gives rise to a transverse charge current
$\vec{j}_{c}^{ISH}(z) = \gamma (2e/\hbar)[\vec{j}_{s,dc}(z) \times
\vec{\sigma}]$ due to the inverse spin Hall effect (ISHE). It has
already been demonstrated that this transverse charge current can be
observed as a {\em dc} voltage
\cite{Saitoh-APL06,Saitoh-PRB08,Azevedo-JAP05}. Here we show how an
approach based on spin pumping can be applied to various F$|$N
combinations.  We identify two contributions to the {\em dc} voltage
that stem from anisotropic magnetoresistance (AMR) and spin Hall
effect, respectively, and can be distinguished by their symmetries.
Furthermore, we present a self-consistent theory that enables
quantification of the spin Hall angle with high accuracy.

\begin{figure}
  \includegraphics[width=8.6cm]{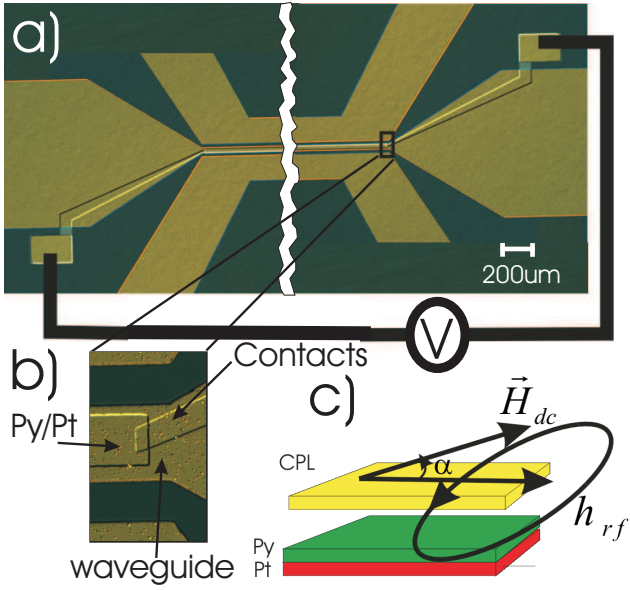}
  \caption{(Color online) Experimental setup.  (a) Optical image of the Py$|$Pt bilayer integrated into the coplanar waveguide. (b) Voltage contacts added at the end of the bilayer for measuring the voltage along the direction of the waveguide.  (c) Directions of the applied {\em dc} magnetic field $\vec{H}_{dc}$ and the {\em rf} driving field $\vec{h}_{rf}$ with respect to the bilayer and waveguide.} %
  \label{setup1}%
\end{figure}

We integrated F$|$N bilayers into coplanar waveguides with
additional leads for measuring a {\em dc} voltage along the sample.
This is shown in Fig.~\ref{setup1} for a Ni$_{80}$Fe$_{20}$
(Py)$|$Pt bilayer, with lateral dimensions of
2.92~mm~$\times$~20~$\mu$m and 15-nm thick individual layers.  The
bilayer was prepared by optical lithography, sputter deposition, and
lift-off on a GaAs substrate. Subsequently we prepared Ag contacts
for the voltage measurements, covered the whole structure with
100-nm thick MgO (for {\em dc} insulation between bilayer and
waveguide), and defined a 30-$\mu$m wide and 200-nm thick Au
coplanar waveguide on top of the bilayer. Similar samples were
prepared with 60-nm thick Au and Mo layers replacing Pt.

\begin{figure}%
  \includegraphics[width=8.6cm, bb=16 15 100 135]{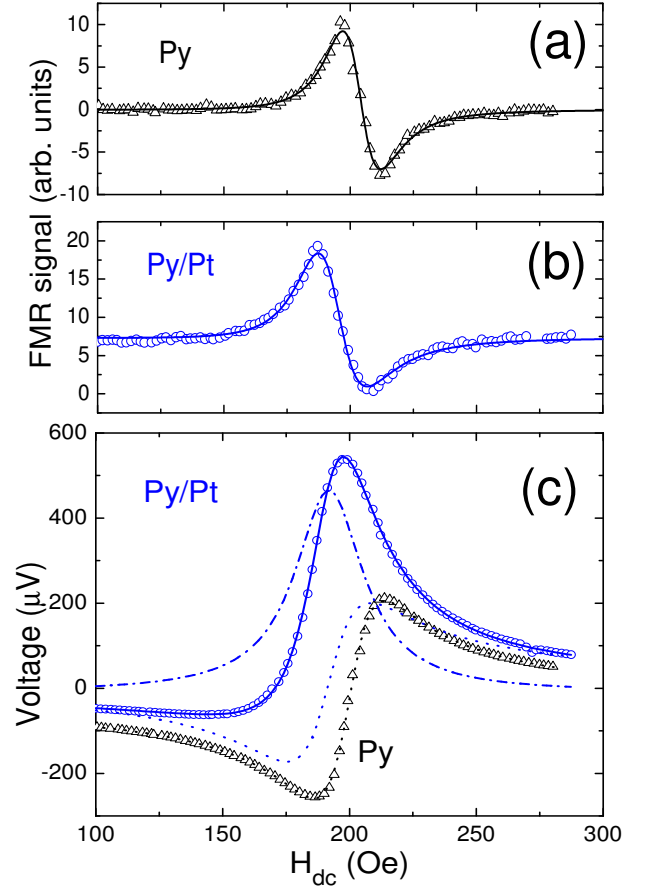}
  \caption{(Color online) (a) and (b) FMR spectra (derivative) for Py$|$Pt (blue $\circ$) and Py (black  $\vartriangle$). Solid lines are fits to a Lorentzian FMR absorption function. (c) Voltage along the samples \textit{vs.} field $H_{dc}$ (Py$|$Pt: blue $\circ$; Py: black  $\vartriangle$).  Dotted and dashed lines are fits to Eqs.~(\ref{AMR}) and (\ref{ISHE}), respectively; the solid line shows the combined fit for the Py$|$Pt sample.} %
  \label{FMRSHE}%
\end{figure}

The FMR was excited by a 4-GHz, 100-mW {\em rf} excitation, while
applying a {\em dc} magnetic field $\vec{H}_{dc}$ at $\alpha =
45^\circ$ with respect to the waveguide [see Fig.~\ref{setup1}(c)].
The FMR signal was determined from the impedance of the waveguide
\cite{Mosendz-JAP08}; simultaneously the {\em dc} voltage was
measured as a function of $\vec{H}_{dc}$. Figure~\ref{FMRSHE} shows
this for a Py$|$Pt bilayer and a Py single layer, where both FMR
peak positions are similar and consistent with the Kittel formula:
\begin{equation}
  \left(\omega / \gamma_{g}\right)^{2}=H_{dc}(H_{dc}+4\pi M_{s}) \label{omega} \;,
\end{equation}
where $\gamma_{g}=g e /2mc$ is the gyromagnetic ratio, $g$ is the
electron g-factor and $M_{s} = 852$~G is the saturation
magnetization for Py. The FMR linewidths (HWHM) extracted from fits
to Lorentzian absorption functions are $\Delta H_{Pt/Py} = 16.9$~Oe
for Py$|$Pt and $\Delta H_{Py} = 12.9$~Oe for Py.  The difference in
FMR linewidth can be attributed to the loss of spin momentum in Py
due to relaxation of the spin accumulation in Pt. This permits the
determination of the additional interface damping due to spin
pumping~\cite{Urban-PRL01}, which in turn provides the interfacial
spin mixing conductance as:
\begin{equation}
  g_{\uparrow\downarrow}=\frac{4\pi\gamma_{g}M_{s}d_{Py}}{g\mu_{B}\omega}(\Delta H_{Pt/Py}-\Delta H_{Py}) \label{spinmix} \;,
\end{equation}
where $d_{Py}$ is the Py layer thickness and $\mu_{B}$ is the Bohr
magneton (spin backflow being disregarded since Pt is an efficient
spin sink). The calculated value for $g_{\uparrow\downarrow} = 2.1
\times10^{19}$~m$^{-2}$ is somewhat smaller than  the previously
reported $2.58 \times 10^{19}$~m$^{-2}$
\cite{Mizukami-JMMM2001,Yaroslav-2002}, but Cao \textit{et\ al.}\
\cite{Cao-JAP09} showed that for high power {\em rf} excitation, the
spin mixing conductance is reduced due to the loss of coherent spin
precession in the ferromagnet.

Figure~\ref{FMRSHE}(c) shows the {\em dc} voltage measured along the
samples. For the Py$|$Pt sample we observe a resonant increase in
the {\em dc} voltage along the sample at the FMR position. However
the lineshape is complicated: below the resonance field the voltage
is negative, it changes sign just before the FMR resonance field,
and has a positive tail in the high field region.  In contrast, the
single layer Py sample, which is not affected by spin pumping, shows
a voltage signal that is purely antisymmetric with respect to the
FMR position.  The voltage due to ISHE depends only on the cone
angle of the magnetization precession [see
Eq.~(\ref{dcspincurrent})] and thus must be symmetric with respect
to the FMR resonance position. This means that the voltage measured
in the Py$|$Pt sample has two contributions: (i) a symmetric signal
due to ISHE and (ii) an antisymmetric signal of the same origin as
in the Py control sample.

\begin{figure}%
  \includegraphics[width=7cm]{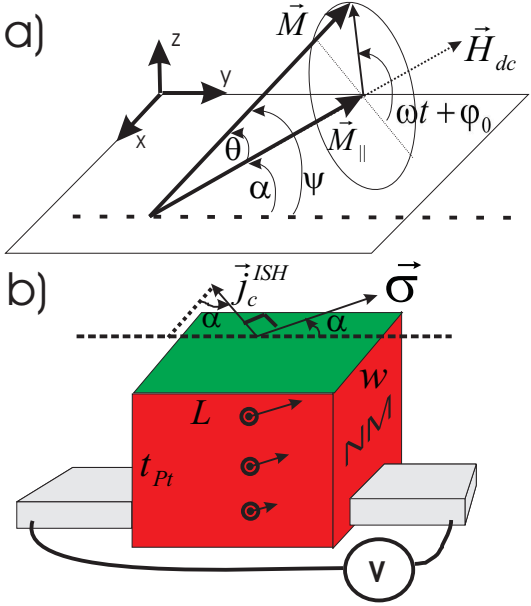}
  \caption{(Color online) (a) Schematic of $\vec{M}$ precessing in Py. $\vec{M}$ precesses around its equilibrium direction given by $\vec{H}_{dc}$ at driving frequency $\omega$ and with phase delay $\varphi_{0}$ with respect to $h_{rf}$.  $\alpha$ is the angle between $\vec{H}_{dc}$ and the waveguide axis (along $y$), $\theta$ is the cone angle described by $\vec{M}$ and $\psi$ is the angle between $\vec{M}$ and the waveguide axis. (b) Geometry of the \textit{dc} component of the pumped spin current with polarization direction $\vec{\sigma}$ along the equilibrium magnetization direction $\vec{M}_{||}$.  The charge current due to ISHE $\vec{j}_{c}^{ISH}$ is orthogonal to the spin current direction (normal to the interface) and $\vec{\sigma}$.  The voltage due to ISHE is measured along $y$ (waveguide axis).  Solid arrows indicate the spin accumulation inside the N, which decays with the spin diffusion length $\lambda_{sd}$.}
  \label{setup2}%
\end{figure}

The antisymmetric voltages observed in both Py and Py$|$Pt originate
from anisotropic magnetoresistance (AMR).  Although the MgO provides
{\em dc} insulation between the sample and the waveguide, there is
strong capacitive coupling, and thus part of the {\em rf} driving
current $I_{rf} = I_{rf}^{m}\sin \omega t$ flows through the sample.
This {\em rf} current in the sample $I_{rf,S}$ flows along the
waveguide direction and its magnitude can be estimated from the
ratio between the waveguide resistance $R_{wg}$ and the sample
resistance $R_S$: $I_{rf,S} = I_{rf}R_{wg}/R_S$. The precessing
magnetization in the Py [see Fig.~\ref{setup2}(a)] results in a
time-dependent $R_S[\psi (t)] = R_{0} - \Delta R_{AMR}\sin^2\psi(t)$
due to AMR given by $\Delta R_{AMR}$, where $R_{0}$ is the sample
resistance with the magnetization along the waveguide axis and
$\psi$ is the angle between the instantaneous magnetization
$\vec{M}$ and the waveguide axis [see Fig.~\ref{setup2}(a)]
\cite{Costache-APL06}. Since the AMR contribution to the resistance
oscillates at the same frequency as the {\em rf} current, a homodyne
{\em dc} voltage develops and is given by:
\begin{equation}\label{AMR}
  V_{AMR}=I_{rf}^{m}\frac{R_{wg}}{R_{S}}
  \Delta R_{AMR}\frac{\sin(2\theta)}{2}\frac{\sin(2\alpha)}{2}\cos\varphi_{0} \; ,
\end{equation}
where $\varphi_{0}$ is the phase angle between magnetization
precession and driving {\em rf} field and the relation between
$\theta$, $\alpha$ and $\psi$ is illustrated in
Fig.~\ref{setup2}(a).  The phase angle $\varphi_0$ is zero well
below the FMR resonance, $\pi /2$ at the peak, and $\pi$ far above
the resonance \cite{Bailey-JMMM06}.  Thus $\cos\varphi_{0}$ changes
sign upon going through the resonance and this gives rise to an
antisymmetric $V_{AMR}$ as observed in both Py and Py$|$Pt samples.
Following Guan \textit{et\ al.}\ \cite{Bailey-JMMM06} we calculate
the cone angle $\theta$ and $\sin\varphi_{0}$ as a function of the
applied field $H_{dc}$, FMR resonance field $H_r$, FMR linewidth
$\Delta H$ and {\em rf} driving field $h_{rf}$:
\begin{equation}
\theta = \frac{h_{rf}\cos\alpha}{\Delta H\sqrt{1+
\left(\frac{(H_{dc}-H_{r})(H_{dc}+H_{r}+4\pi M_{s})}{\Delta H4\pi
M_{s}}\right)^{2}}} \label{teta}, \mathrm{and}
\end{equation}
\begin{equation}
\sin\varphi_{0} = \frac{1}{\sqrt{1+\left(\frac{(H_{dc}-H_{r})(H_{dc}+H_{res}+4\pi M_{s})}{\Delta H4\pi
M_{s}}\right)^{2}}} \label{phi0} \;.
\end{equation}
Using Eqs.~(\ref{AMR}--\ref{phi0}) and taking a measured 0.95\%
value for $\Delta R_{AMR}$ fits the Py data [see
Fig.~\ref{FMRSHE}(b)] with only one adjustable parameter
$h_{rf}=4.5$~Oe.

In order to understand the Py$|$Pt voltage data we have to include
an additional contribution due to ISHE.  In an open circuit an
electric field $\vec{E}$ is generated leading to a total current
density $\vec{j}(z)=\vec{j}_{c}^{ISH}(z)+\sigma_{N}\vec{E}$ with
$\int\vec{j}(z)dz=0$ where $\sigma_{N}$ is the N conductivity. When
the wire is much longer than thick, the electric field is constant
in the wire and the component of the electric field along the
measurement direction $y$ is:
\begin{equation}\label{efield}
E_{y}=-\frac{\gamma}{\sigma_{N}}\frac{e\omega}{2\pi}g_{\uparrow\downarrow}\sin\alpha\sin^{2}\theta\frac{\lambda_{sd}}{t_{N}}
\tanh\left(\frac{t_{N}}{2\lambda_{sd}}\right).
\end{equation}
Using Eq.~\ref{efield} we calculate the voltage due to ISHE
generated along the sample with length $L$:
\begin{equation}\label{ISHE}
  V_{ISH}=-\frac{\gamma g_{\uparrow\downarrow} eL \lambda_{sd} \omega}{2\pi \sigma_{N} t_{N}}
  \sin\alpha\sin^{2}\theta\tanh\left(\frac{t_{N}}{2\lambda_{sd}}\right) \; .
\end{equation}
Note that this voltage is proportional to $L$ and thus measurements
of small $\gamma$ can be achieved by increasing the sample
dimension.  We used Eqs.~(\ref{ISHE}) and~(\ref{AMR}) to fit the
voltage measured for the Py$|$Pt sample, see solid line in
Fig.~\ref{FMRSHE}(c).  The dashed and dotted lines in
Fig.~\ref{FMRSHE}(c) are the AMR and ISHE contributions,
respectively.  By using a literature value for Pt of $\lambda_{sd} =
10 \pm 2$~nm \cite{Kurt-APL02}, the only remaining adjustable
parameters are the {\em rf} driving field $h_{rf} = 4.5$~Oe and the
spin Hall angle $\gamma = 0.0067 \pm 0.0006$. Note that through the
cone angle $\theta$, $h_{rf}$ enters both the AMR and ISHE
contributions; this puts an additional constraint on this parameter,
and, in fact, as seen from the fit to the control Py sample, it is
already determined by the negative and positive tails of the AMR
part.

\begin{figure}%
  \includegraphics[width=8.6cm,bb=16 15 100 80]{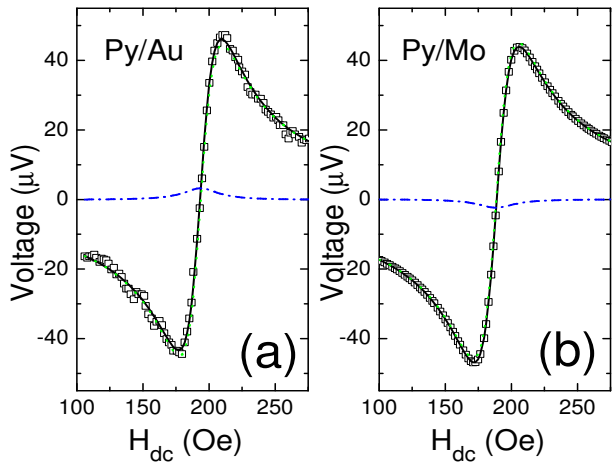}
  \caption{(Color online) Voltage measured for (a) Py$|$Au and (b) Py$|$Mo.  Shown are data (symbols), combined fits (solid lines) and individual AMR and ISHE contributions, with dotted (overlapping with solid) and dashed lines, respectiely.   Note the opposite sign of ISHE contributions for Au and Mo.} %
    \label{MoAu}%
\end{figure}

\begin{table}
\caption{Spin Hall angle $\gamma$ determined using $\lambda_{sd}$
and $\sigma_{N}$}
\begin{ruledtabular}
\begin{tabular}{cccr}\label{gammas}
Normal metal & $\lambda_{sd}$ (nm) & $\sigma_{N}$ $1/(\Omega m)$&\multicolumn{1}{c}{$\gamma$} \\
\hline
Pt   & 10$\pm$2 & (2.42$\pm$0.19)$\times10^{6}$ & 0.0067$\pm$0.0006 \\
Au  &  35$\pm$3 & (2.52$\pm$0.13)$\times10^{7}$ &  0.0016$\pm$0.0003 \\
Mo  & 35$\pm$3 & (4.66$\pm$0.23)$\times10^{6}$ & -0.00023$\pm$0.00005 \\
\end{tabular}
\end{ruledtabular}
\end{table}

This technique can be readily applied to determine $\gamma$ in any
conductor. In Fig.~\ref{MoAu} we show voltages measured for Py$|$Au
and Py$|$Mo. The spin Hall contributions in Au and Mo are smaller
than in Pt, and note that for Mo the spin Hall contribution changes
its sign. Fitting of the data enabled us to extract the values of
$\gamma$ for Au and Mo, see Table~\ref{gammas}. Note that the
determination of $\gamma$ requires $\sigma_{N}$ and $\lambda_{sd}$
as an input parameters. $\sigma_{N}$ was measured using four-probe
measurements for all samples. Reported values for $\lambda_{sd}$
vary considerably. We choose a conservatively low literature value
for Pt from Ref.~\onlinecite{Kurt-APL02} and Au from
Ref.~\onlinecite{Mosendz-PRB09}, and for Mo we assumed that
$\lambda_{sd}$ is comparable to Au. Even though this latter
assumption may not necessarily hold, the sign change is consistent
with earlier measurements~\cite{Morota-JAP09}. Furthermore, our
observed values for $\gamma$ are in good agreement with values
reported by Otani \textit{et\ al.}\ \cite{Vila-PRL07,Morota-JAP09}
from measurements in lateral spin valves, but conflict with more
optimistic values reported by other groups
\cite{Seki-NM2008,Ando-PRL2008}.  We note that in lateral spin
valves it is important to also understand the charge current
contribution in order to rule out additional non-local voltage
contributions \cite{Mihajlovic-Preprint}.  In contrast, in our
approach the spin pumping creates a uniform, macroscopic and
well-defined spin current across the whole sample, and the voltage
signal from spin Hall effects can readily be increased through use
of longer samples, since $V_{ISH} \propto L$.  Furthermore, using an
integrated coplanar waveguide architecture provides control over
parameters, such as the {\em rf} driving field distribution.  This
enables us to carry out a quantitative analysis of the data, in
contrast to the more qualitative description of the ISHE in
Refs.~\onlinecite{Saitoh-APL06} and~\onlinecite{Saitoh-PRB08}.

In conclusion, we performed FMR with simultaneous transverse voltage
measurements in ferromagnetic/normal metal bilayers.  From this we
accurately determine the spin Hall angle for Pt, Au and Mo by
fitting the experimental data to a theory, which accounts for both
the anistropic magnetoresistance and inverse spin Hall effect
contributions. The combination of spin pumping and spin Hall effects
provides a valuable technique for measuring spin Hall angle in many
different materials.

We would like to thank R. Winkler, G. Mihajlovi\'{c} and M. Dyakonov
for valuable discussions. This work was supported by U.S. DOE-BES
under Contract No.\ DE-AC02-06CH11357.


\begin{thebibliography}{28}
\bibitem{endnote}
  \bibinfo{journal}{circular precession
was assumed for simplicity, realistic elliptical precession
trajectory will result in a quantitative, but not qualitative
modification to the model}.
\end{thebibliography}

\end{document}